\documentclass[a4paper]{report}

\usepackage{RJournal}
\usepackage[round]{natbib}
\usepackage{setspace}
\usepackage{natbib}
\usepackage{xspace}
\usepackage[utf8]{inputenc}
\usepackage{pmboxdraw}

\begin{document}

\fancyhf{}
\fancyhead[LO,RE]{\textsc{Contributed Article}}
\fancyhead[RO,LE]{\thepage}
\fancyfoot[L]{The R Journal Vol. 5/1, June 2013}
\fancyfoot[R]{ISSN 2073-4859}


\begin{article}
\title{Possible Directions for Improving Dependency Versioning in R}
\author{by Jeroen Ooms}

\maketitle

\abstract{
One of the most powerful features of R is its infrastructure for contributed
code. The built-in package manager and complementary repositories
provide a great system for development and exchange of code, and have played an
important role in the growth of the platform towards the de-facto standard in
statistical computing that it is today. However, the number of packages on CRAN
and other repositories has increased beyond what might have been foreseen, and
is revealing some limitations of the current design. One such problem is the
general lack of dependency versioning in the infrastructure. This paper explores
this problem in greater detail, and suggests approaches taken by other open
source communities that might work for R as well. Three use cases are defined that
exemplify the issue, and illustrate how improving this aspect of package management
could increase reliability while supporting further growth of the R community.}

\section{Package management in R}

One of the most powerful features of R is its infrastructure for contributed
code \citep{fox2009aspects}. The base R software suite that is released several
times per year ships with the \dfn{base} and \dfn{recommended} packages and
provides a solid foundation for statistical computing. However, most R users
will quickly resort to the package manager and install packages contributed by
other users. By default, these packages are installed from the ``Comprehensive
R Archive Network'' (\CRANpkg{CRAN}), featuring over 4300 contributed packages as
of 2013. In addition, other repositories like BioConductor \citep{bioc} and
Github \citep{dabbish2012social} are hosting a respectable number of packages
as well.

The \emph{R Core team} has done a tremendous job in coordinating the
development of the base software along with providing, supporting, and maintaining an
infrastructure for contributed code. The system for sharing and installing
contributed packages is easily taken for granted, but could in fact not
survive without the commitment and daily efforts from the repository
maintainers. The process from submission to publication of a package involves
several manual steps needed to ensure that all published packages
meet standards and work as expected, on a variety of platforms, architectures
and R versions. In spite of rapid growth and limited resources, CRAN has
managed to maintain high standards on the quality of packages. Before
continuing, we want to express appreciation for the countless hours invested
by volunteers in organizing this unique forum for statistical software.
They facilitate the innovation and collaboration in our field,
and unite the community in creating software that is both of the highest
quality and publicly available. We want to emphasize that suggestions made in
this paper are in no way intended as criticism on the status quo. If anything,
we hope that our ideas help address some challenges to support further growth
without having to compromise on the open and dynamic nature of the
infrastructure.

\subsection{The dependency network}

Most R packages depend on one or more other packages, resulting in a complex
network of recursive dependencies. Each package includes a \file{DESCRIPTION}
file which allows for declaration of several types of dependencies, including
\code{Depends}, \code{Imports}, \code{Suggests} and \code{Enhances}. Based on
the type of dependency relationship, other packages are automatically
installed, loaded and/or attached with the requested package. Package
management is also related to the issue of \dfn{namespacing}, because different
packages can use identical names for objects. The \file{NAMESPACE} file allows
the developer to explicitly define objects to be exported or imported from
other packages. This prevents the need to attach all dependencies and lookup
variables at runtime, and thereby decreases chances of masking and
naming-conflicts. Unfortunately, many packages are not taking advantage of this
feature, and thereby force R to attach all dependencies, unnecessarily filling
the search path of a session with packages that the user hasn't asked for.
However, this is not the primary focus of this paper.

\subsection{Package versioning}

Even though CRAN consistently archives older versions of every package when
updates are published, the R software itself takes limited advantage of this
archive. The package manager identifies packages by name only when installing
or loading a package. The \code{install.packages} function downloads and
installs the \emph{current} version of a CRAN package into a single global
library. This library contains a single version of each package. If a previous
version of the package is already installed on the system, it is overwritten
without warning. Similarly, the \code{library} function will load the earliest
found package with a matching name.

The \file{DESCRIPTION} file does allow the package author to specify a certain
version of a dependency by postfixing the package name with \texttt{>=}, \texttt{<=} or
\texttt{==} and a version string. However, using this feature is actually dangerous
because R might not be able to satisfy these conditions, causing errors. This is
again the result of R libraries, sessions and repositories being limited to a
single current version of each package. When a package would require a version
of a dependency that is not already installed or current on CRAN, it can not be
resolved automatically. Furthermore, upgrading a package in the global library
to the current CRAN version might break other packages that require the
previously installed version. Experienced R users might try to avoid such
problems by manually maintaining separate libraries for different tasks and
projects. However, R can still not have multiple versions of a package loaded
concurrently. This is perhaps the most fundamental problem because it is nearly
impossible to work around. If package authors would actually declare specific
versions of dependencies, any two packages requiring different versions of
one and the same dependency will conflict and cannot be used together. In
practice, this limitation discourages package authors to be explicit about
dependency versions. The \texttt{>=} operator is used by some packages, but
it only checks if an installed dependency is outdated and needs to be
synchronized with CRAN. It still assumes that any current of future version will
suffice, and does not protect packages from breaking when their dependency
packages change. The \texttt{<=} and \texttt{==} operators are barely used at
all.

When identifying a package by its name only, we implicitly make the assumption
that different versions of the package are interchangeable. This basic
assumption has far-reaching implications and consequences on the distributed
development process and reliability of the software as a whole. In the context
of the increasingly large pool of inter-dependent packages, violations of this
assumption are becoming increasingly apparent and problematic. In this paper we
explore this problem is greater detail, and try to make a case for moving away
from this assumption, towards systematic versioning of dependency relationships.
The term \dfn{dependency} in this context does not exclusively refer to formally
defined relations between R packages. Our interpretation is a bit more general in
the sense that any R script, Sweave document, or third party application
\dfn{depends} on R and certain packages that are needed to make it function.
The paper is largely motivated by personal experiences, as we have come to
believe that limitations of the current dependency system are underlying
multiple problems that R users and developers might experience. Properly
addressing these concerns could resolve several lingering issues at once,
and make R a more reliable and widely applicable analytical engine.

\section{Use cases}

A dependency defines a relationship wherein a certain piece of software requires
some other software to run or compile. However, software constantly evolves, and in
the open source world this happens largely unmanaged. Consequently, any software
library might actually be something different today than it was yesterday.
Hence, solely defining the dependency relationship in terms of the name of the
software is often insufficient. We need to be more specific, and declare
explicitly which version(s), branch(es) or release(s) of the other software
package will make our program work. This is what we will refer to as
\dfn{depencency versioning}.

This problem is not at all unique to R; in fact a large share of this paper
consist of taking a closer look at how other open source communities are
managing this process, and if some of their solutions could apply to R as
well. But first we will elaborate a bit further on how this problem exactly
appears in the context of R. This section describes three use cases that
reveal some limitations of the current system. These use cases delineate
the problem and lead towards suggestions for improvements in subsequent sections.

\subsection{Case 1: Archive / repository maintenance}

A medium to large sized repository with thousands of packages has a complicated
network of dependencies between packages. CRAN is designed to consider the very
latest version of every package as the only \dfn{current} version. This design
relies on the assumption that at any given time, the latest versions
of all packages are compatible. Therefore, R's built-in package manager can simply
download and install the current versions of all dependencies along with the
requested package, which seems convenient. However, to developers this means
that every package  \dfn{update} needs to maintain full backward compatibility
with all previous versions. No version can introduce any breaking changes, because
other packages in the repository might be relying on things in a certain way.
Functions or objects may never be removed or modified; names, arguments, behavior,
etc, must remain the same. As the dependency network gets larger and more complex,
this policy becomes increasingly vulnerable. It puts a heavy burden on
contributing developers, especially the popular ones, and results in
increasingly large packages that are never allowed to deprecate or clean up
old code and functionality.

In practice, the assumption is easily violated. Every time a package update is
pushed to CRAN, there is a real chance of some reverse dependencies failing due
to a breaking change. In the case of the most popular packages, the probability
of this happening is often closer to 1 than to 0, regardless of the author. Uwe
Ligges has stated in his keynote presentation at useR that CRAN automatically
detects some of these problems by rebuilding every package up in the dependency
tree. However, only a small fraction of potential problems reveal themselves
during the build of a package, and when found, there is no obvious solution. One
recent example was the forced roll-back of the \pkg{ggplot2} \citep{ggplot2}
update to version 0.9.0, because the introduced changes caused several other
packages to break. The author of the \pkg{ggplot2} package has since been
required to announce upcoming updates to authors of packages that depend on
\pkg{ggplot2}, and provide a release candidate to test compatibility. The
dependent packages are then required to synchronize their releases if any
problems arise. However, such manual solutions are far from flawless and put
even more work on the shoulders of contributing developers. It is doubtful that
all package authors on CRAN have time and resources to engage in an extensive
dialogue with other maintainers for each update of a package. We feel strongly
that a more systematic solution is needed to guarantee that software published
on CRAN keeps working over time; current as well as older versions.

When the repository reaches a critical size, and some packages collect hundreds
of reverse dependencies, we have little choice but to acknowledge the fact that
every package has only been developed for, and tested with certain versions of
its dependencies. A policy of assuming that any current or future version of a
dependency should suffice is dangerous and sets the wrong incentives for package
authors. It discourages change, refactoring or cleanup, and results in packages
accumulating an increasingly heavy body of legacy code. And as the repository
grows, it is inevitable that packages will nevertheless eventually break as
part of the process.
What is needed is a redesign that supports the continuous decentralized change
of software and helps facilitate more reliable package development. This is not
impossible: there are numerous open source communities managing repositories with
more complex dependency structures than CRAN. Although specifics vary, they form
interesting role models to our community. As we will see later on, a properly
archived repository can actually come to be a great asset rather than a
liability to the developer.

\subsection{Case 2: Reproducibility}

Replication is the ultimate standard by which scientific claims are judged. However,
complexity of data and methods can make this difficult to achieve computational
science \citep{peng2011reproducible}. As a leader in scientific computing, R
takes a pioneering role in providing a system that encourages researchers to strive
towards the gold standard. The CRAN Task View on Reproducible Research states
that:

\begin{quote}
\emph{The goal of reproducible research is to tie specific instructions
to data analysis and experimental data so that scholarship can be recreated,
better understood and verified.}
\end{quote}
In R, reproducible research is largely facilitated using literate programming
techniques implemented in packages like \pkg{Sweave} that mix (weave) R code
with \LaTeX-markup to create a ``reproducible document''
\citep{leisch2002sweave}. However, those ever faced with the task of actually
reproducing such a document might have experienced that the Sweave file does not
always compile out of the box. Especially if it was written several years ago
and loads some contributed packages, chances are that essential things have
changed in the software since the document was created. When we find ourselves
in such a situation, recovering the packages needed to reproduce the document
might turn out to be non-trivial.

An example: suppose we would like to reproduce a Sweave document which was
created with R 2.13 and loads the \pkg{caret} package \citep{caret}. If no further
instructions are provided, this means that any of the approximately 25 releases
of \pkg{caret} in the life cycle of R 2.13 (April 2011 to February 2012) could
have been used, making reproducibility unlikely. Sometimes authors add comments
in the code where the package is loaded, stating that e.g. \pkg{caret 4.78} was
used. However, this information might also turn out to be insufficient:
\pkg{caret} depends on 4 packages, and suggests another 59 packages, almost all of which
have had numerous releases in R 2.13 time frame.
Consequently, \pkg{caret 4.78} might not work anymore because of changes in
these dependencies. We then need to do further investigation to figure out
which versions of the dependency packages were current at the time of the
\pkg{caret 4.78} release. Instead, lets assume that the prescient researcher
anticipated all of this, and saved the full output of \code{sessionInfo()}
along with the Sweave document, directly after it was compiled. This output
lists the version of each loaded package in the active R session.
We could then proceed by manually downloading and installing R 2.13 along with
all of the required packages from the archive. However, users on a commercial
operating systems might be up for another surprise: unlike source packages,
binary packages are not fully archived. For example, the only binary builds
available for R 2.13 are respectively \pkg{caret 5.13} on Windows, and
\pkg{caret 5.14} on OSX. Most likely, they will face the task of rebuilding
each of the required packages from source in an attempt to reconstruct the
environment of the author.

Needless to say, this situation is suboptimal. For manually compiling a single
Sweave document we might be willing to make this effort, but it does not
provide a solid foundation for systematic or automated reproducible software
practices. To make results generated by R more reproducible, we need better
conventions and/or native support that is both explicit and specific about
contributed code. For an R script or Sweave document to stand the test of time,
it should work at least on the same version of R that was used by the author. In
this respect, R has higher requirements on versioning than other software.
Reproducible research does not just require a version that will
make things work, but one that generates exactly the same output. In
order to systematically reproduce results R, package versions either need to be
standardized, or become a natural part of the language. We realize this will
not archive perfect reproducibility, as problems can still arise due to OS or
compiler specific behavior. However, it will be a major step forward that has
the potential of turning reproducibility into a natural feature of the
software, rather than a tedious exercise.
 
\subsection{Case 3: Production applications}

R is no longer exclusively used by the local statistician through an
interactive console. It is increasingly powering systems, stacks and
applications with embedded analytics and graphics. When R is part of say, an
application used in hospitals to create on-demand graphics from patient data,
the underlying code needs to be stable, reliable, and redistributable. Within
such an application, even a minor change in code or behavior can result in
complete failure of the system and cannot easily be fixed or debugged.
Therefore, when an application is put in production, software has to be
completely frozen.

An application that builds on R has been developed and tested with certain
versions of the base software and R packages used by the application. In order
to put this application in production, exactly these versions need to be shipped,
installed and loaded by the application on production servers. Managing,
distributing and deploying production software with R is remarkably hard, due
to limited native dependency versioning and the single global library design.
Administrators might discover that an application that was working in one
place doesn't work elsewhere, even though exactly the same operating system,
version of R, and installation scripts were used. The problem of course is that
the contributed packages constantly change. Problems become more complicated
when a machine is hosting many applications that were developed by different
people and depend on various packages and package versions.

The default behavior of loading packages from a global library with bleeding
edge versions is unsuitable for building applications. Because the CRAN
repository has no notion of stable branches, one manually needs to download and
install the correct versions of packages in a separate library for each
application to avoid conflicts. This is quite tricky and hard to scale when
hosting many applications. In practice, application developers might not even be
aware of these pitfalls, and design their applications to rely on the default
behavior of the package manager. They then find out the hard way that
applications start breaking down later on, because of upstream changes or
library conflicts with other applications.

\section{Solution 1: staged distributions}

The problem of managing bottom-up decentralized software development is not
new; rather it is a typical feature of the open source development
process. The remainder of this paper will explore two solutions from other open
source communities, and suggest how these might apply to R. The current section
describes the more classic solution that relies on staged software
\dfn{distributions}.

A \dfn{software distribution} (also referred to as a \dfn{distribution} or a \dfn{distro})
is a collection of software components built, assembled and configured so that
it can be used essentially "as is" for its intended purpose. Maintainers of
distributions do not develop software themselves; they collect software from
various sources, package it up and redistribute it as a system. Distributions
introduce a formal release cycle on the continuously changing upstream
developments and maintainers of a distribution take responsibility for ensuring
compatibility of different packages within a certain release of the
distribution. Software distributions are most commonly known in the context of
free operating systems (BSD, Linux, etc). Staging and shipping software in a
distribution has proven to scale well to very large code bases. For example,
the popular Debian GNU/Linux distribution (after which R's package description
format was modeled) features over 29000 packages with a large and complex
dependency network. No single person is familiar with even a fraction of the
code base that is hosted in this repository. Yet through well organized staging and
testing, this distribution is known to be one of the most reliable operating
systems today, and is the foundation for a large share of the global IT
infrastructure.

\subsection{The release cycle}

In a nutshell, a staged distribution release can be organized as follows. At any
time, package authors can upload new versions of packages to the \dfn{devel}
pool, also known as the \dfn{unstable} branch. A release cycle starts with
distribution maintainers announcing a \dfn{code freeze} date, several
months in advance. At this point, package authors are notified to ensure
that their packages in the unstable branch are up to date, fix bugs and resolve
other problems. At the date of the code freeze, a copy (fork) of the unstable
repository is made, named and versioned, which goes into the \dfn{testing}
phase. Software in this branch will then be subject to several iterations of
intensive testing and bug fixing, sometimes accompanied by \dfn{alpha} or
\dfn{beta} releases of the distribution. However, software versions in the testing branch will
no longer receive any major updates that could potentially have side effects
or break other packages. The goal is to converge to increasingly stable set of
software. When after several testing rounds the distribution maintainers are
confident that all serious problems are fixed, the branch is tagged \dfn{stable}
and released to the public. Software in a stable release will usually only
receive minor non-breaking updates, like important compatibility fixes and
security updates. For the next ``major release'' of any software, the user will
have to wait for the next cycle of the distribution. As such, everyone using a
certain release of the distribution is using exactly the same versions of all programs
and libraries on the system. This is convenient for both users and developers
and gives distributions a key role in bringing decentralized open source
development efforts together.

\subsection{R: downstream staging and repackaging}

The semi annual releases of the r-base software suite can already be considered
as a distribution of the 29 \code{base} and \code{recommended} packages.
However in the case of R, this collection is limited to software that has been centrally
developed and released by the same group of people; it does not include contributed
code. Due to the lack of native support for dependency versioning in
R, several third party projects have introduced some form of downstream staging in order to
create stable, redistributable collections R software. This section lists some
examples and explains why this is suboptimal. In the next section we will
discuss what would be involved with extending the R release cycle to contributed
packages.

One way of staging R packages downstream is by including them in existing
software distributions. For example, \citet{cran2deb} have wrapped some popular
CRAN packages into \texttt{deb} packages for the Debian and Ubuntu systems.
Thereby, pre-compiled binaries are shipped in the distribution along with the R
base software, putting version compatibility in the hands of the maintainers
(among other benefits).
This works well, but requires a lot of effort and commitment from the package
maintainer, which is why this has only been done for a small subset of the CRAN
packages. Most distributions expect high standards on the quality of the
software and package maintenance, which makes this approach hard to scale up to
many more packages. Furthermore, we are tied to the release cycle of the
distribution, resulting in a somewhat arbitrary and perhaps unfortunate
snapshot of CRAN packages when the distribution freezes. Also, different
distributions will have different policies on if, when and which packages they
wish to ship with their system.

Another approach is illustrated by domain-specific projects like BioConductor
(genomic data) and REvolution R Enterprise (big data). Both these systems
combine a fixed version of R with a custom library of frozen R packages. In
the case of REvolution, the full library is included with the installer; for
BioConductor they are provided through a dedicated repository. In both cases,
this effectively prevents installed software from being altered unexpectedly by
upstream changes. However, this also leads to a split in the community between
users of R, BioConductor, and REvolution Enterprise. Because of the differences
in libraries, R code is not automatically portable between these systems,
leading to fragmentation and duplication of efforts. E.g. BioConductor seems to
host many packages that could be more generally useful; yet they are unknown to
most users of R. Furthermore, both projects only target a limited set of
packages; they still rely on CRAN for the majority of the contributed code.

The goal of staging is to tie a fixed set of contributed packages to a certain
release of R. If these decisions are passed down to distributions or organizations, 
a multitude of local conventions and repositories arises, and different
groups of users will still be using different package versions. This leads to
unnecessary fragmentation of the community by system, organization, or distribution
channel. Moreover, it is often hard to assess compatibility of third party packages, 
resulting in somewhat arbitrary local decision making. It seems that the people who
are in the best position to manage and control compatibility are the package authors
themselves. This leads us to conclude that a more appropriate place to organize staging
of R packages is further upstream.

\subsection{Branching and staging in CRAN itself}

Given that the community of R contributors evolves mainly around CRAN, the most
desirable approach to organizing staging would be by integrating it with the
publication process. Currently, CRAN is managed as what distributions would
consider a \dfn{development} or \dfn{unstable} branch. It consists of the pool
of \dfn{bleeding-edge} versions, straight from package authors. Consequently it
is wise to assume that software in this branch might break on a regular
basis. Usually, the main purpose of an
\dfn{unstable} branch is for developers to exchange new versions and test
compatibility of software. Regular users obtain software releases from
\dfn{stable} branches instead. This does not sound unfamiliar: the r-base
software also distinguishes between stable versions \dfn{r-release} and
\dfn{r-release-old}, and an unstable development version, \dfn{r-devel}.

The fact that R already has an semi-annual release cycle for the 29 \code{base}
and \code{recommended} packages, would make it relatively straightforward to
extend this cycle to CRAN packages. A snapshot of CRAN could be frozen along
with every version of \dfn{r-release}, and new package updates would only be
published to the \dfn{r-devel} branch. In practice, this could perhaps quite
easily be implemented by creating a directory on CRAN for each release of R,
containing symbolic links to the versions of the packages considered
\dfn{stable} for this release. In the case of binary packages for OSX and
Windows, CRAN actually already has separate directories with builds for each
release of R. However currently these are not frozen and continuously updated.
In a staged repository, newly submitted packages are only build for the current
\dfn{devel} and \dfn{testing} branches; they should not affect \dfn{stable}
releases. Exceptions to this process could still be granted to authors that need
to push an important update or bugfix within a stable branch, commonly referred
to as \dfn{backporting}, but this should only happen incidentally.

To fully make the transition to a staged CRAN, the default behavior of the
package manager must be modified to download packages from the stable branch of
the current version of R, rather than the latest development release. As such,
all users on a given version of R will be using the same version of each CRAN
package, regardless on when it was installed. The user could still be given an
option to try and install the development version from the unstable branch, for
example by adding an additional parameter to \code{install.packages} named
\code{devel=TRUE}. However when installing an unstable package, it must be
flagged, and the user must be warned that this version is not properly tested
and might not be working as expected. Furthermore, when loading this package a
warning could be shown with the version number so that it is also obvious from
the output that results were produced using a non-standard version of the
contributed package. Finally, users that would always like to use the very
latest versions of all packages, e.g. developers, could install the
\texttt{r-devel} release of R. This version contains the latest commits by R
Core and downloads packages from the devel branch on CRAN, but should not be
used or in production or reproducible research settings.

\subsection{Organizational change}

Appropriate default behavior of the software is a key element to encourage
adoption of conventions and standards in the community. But just as important is
communication and coordination between repository maintainers and package authors.
To make staging work, package authors must be notified of upcoming deadlines,
code freezes or currently broken packages. Everyone must realize that the
package version that is current at the time of code freeze, will be used by the
majority of users of the upcoming version of R. Updates to already released
\emph{stable} branches can only be granted in exceptional circumstances, and
must guarantee to maintain full backward compatibility. The policies of the
BioConductor project provide a good starting point and could be adapted to work
for CRAN.

Transitioning to a system of ``stable'' and ``development'' branches in CRAN,
where the stable branch is conventional for regular users, could tremendously
improve the reliability of the software. The version of the R software itself
would automatically imply certain versions of contributed packages. Hence, all
that is required to reproduce a Sweave document created several years ago, is
which version of R was used to create the document. When deploying an
application that depends on R 2.15.2 and various contributed packages, we can
be sure that a year later the application can be deployed just as easily, even
though the authors of contributed packages used by the application might have
decided to implement some breaking changes.
And package updates that deprecate old functionality or might break other
packages that depend on it, can be uploaded to the \dfn{unstable} branch
without worries, as the stable branches will remain unchanged and users won't
be affected. The authors of the dependent packages that broke due to the update
can be warned and will have sufficient time to fix problems before the next
\dfn{stable} release.

\section{Solution 2: versioned package management}

The previous section described the ``classical'' solution of creating distributable
sets of compatible, stable software. This is a proven approach and has been
adopted in some way or another by many open-source communities. However, one
drawback of this approach might be that some additional coordination is needed
for every release. Another drawback is that it makes the software a bit
more conservative, in the sense that regular users will generally be using
versions of packages that are at least a couple of months old. The current
section describes a different approach to the problem that is used by for
example the Javascript community. This method is both reliable and flexible,
however would require some more fundamental changes to be implemented in R.

\subsection{Node.js and NPM}

One of the most recent and fastest growing open source communities is that of
the node.js software (for short: \dfn{node}), a Javascript server system based
on the open source engine \dfn{V8} from Google. One of the reasons that the
community has been able to grow rapidly is because of the excellent package
manager and identically named repository, \dfn{NPM}. Even though this package
manager is only 3 years old, it is currently hosting over 30000 packages with
more than a million downloads daily, and has quickly become the standard
way of distributing Javascript code. The NPM package manager is a powerful tool
for development, publication and deployment of both libraries and applications.
NPM addresses some problems that Javascript and R actually have in common, and
makes an interesting role model for a modern solution to the problem.\\

\begin{figure}[htbp]
  \centering
  \includegraphics[width=0.3\textwidth]{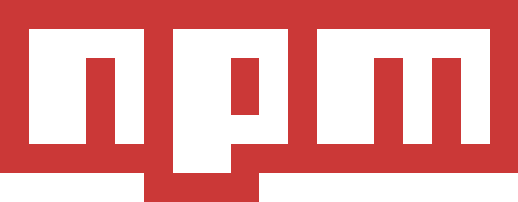}
\end{figure}

\noindent The Javascript community can be described as decentralized,
unorganized and highly fragmented development without any quality control authority. Similar to
CRAN, NPM basically allows anyone to claim a ``package name'' and start
publishing packages and updates to the repositories. The repository has no
notion of branches and simply stores every version of a package indefinitely in
its archives. However, a major difference with R is how the package manager
handles installation, loading and namespacing of packages.

\subsection{Dependencies in NPM}

Every \dfn{NPM} package ships with a file named \file{package.json}, which is
the equivalent of the \file{DESCRIPTION} in R packages, yet a bit more advanced.
An overview of the full feature set of the package manager is beyond the scope
of this paper, but the interested reader is highly encouraged to take a look
over the fence at this well designed system: \url{https://npmjs.org/doc/json.html}.
The most relevant feature in the context CRAN is how NPM declares and resolves
dependencies.

Package dependencies are defined using a combination of the package \emph{name}
and \emph{version range descriptor}. This descriptor is specified with a
simple dedicated syntax, that extends some of the standard versioning notation.
Below a snippet taken from the \file{package.json} file in the NPM manual:

\begin{verbatim}
   "dependencies" : {
      "foo" : "1.0.0 - 2.9999.9999",
      "bar" : ">=1.0.2 <2.1.2",
      "baz" : ">1.0.2 <=2.3.4",
      "boo" : "2.0.1",
      "qux" : "<1.0.0 || >=2.3.1 <2.4.5",
      "asd" : "http://asdf.com/asdf.tar.gz",
      "til" : "~1.2",
      "elf" : "~1.2.3",
      "two" : "2.x",
      "thr" : "3.3.x",
   }
\end{verbatim}

\noindent The version range descriptor syntax is a powerful tool to specify
which version(s) or version range(s) of dependencies are required. It provides the exact
information needed to build, install and/or load the software. In contrast to R,
NPM takes full advantage of this information. In R, all packages are installed
in one or more global libraries, and at any given time a subset of these packages
is loaded in memory. This is where NPM takes a very different approach. During
installation of a package, NPM creates a \emph{subdirectory} for dependencies inside
the installation directory of the package. It compares the list of dependency
declarations from the \file{package.json} with an index of the repository archive,
and then constructs a private library containing the full dependency tree and precise
versions as specified by the author. Hence, every installed package has its own library
of dependencies. This works recursively, i.e. every dependency package inside
the library again has its own dependency library.

\begin{verbatim}
jeroen@ubuntu:~/Desktop$ npm install d3
jeroen@ubuntu:~/Desktop$ npm list
/home/jeroen/Desktop
└─┬ d3@2.10.3
  ├─┬ jsdom@0.2.14
  │ ├─┬ contextify@0.1.3
  │ │ └── bindings@1.0.0
  │ ├── cssom@0.2.5
  │ ├── htmlparser@1.7.6
  │ └─┬ request@2.12.0
  │   ├─┬ form-data@0.0.3
  │   │ ├── async@0.1.9
  │   │ └─┬ combined-stream@0.0.3
  │   │   └── delayed-stream@0.0.5
  │   └── mime@1.2.7
  └── sizzle@1.1.0
\end{verbatim}

\noindent By default, a package loads dependencies from its private library, and
the namespace of the dependency is imported explicitly in the code. This way, an
installed NPM package is completely unaffected by other applications, packages,
and package updates being installed on the machine. The private library of any
package contains all required dependencies, with the exact versions that
were used to develop the package. A package or application that has been
tested to work with certain versions of its dependencies, can easily be
installed years later on another machine, even though the latest versions of
dependencies have had major changes in the mean time.

\subsection{Back to R}

A similar way of managing packages could be very beneficial to R as well. It would
enable the same dynamic development and stable installation of packages that has
resulted in a small revolution within the Javascript community. The only serious
drawback of this design is that it requires more disk space and slightly
more memory, due to multiple versions packages being installed and/or loaded.
Yet the memory required to load an additional package is minor in comparison
with loading and manipulating a medium sized dataset. Considering the wide
availability of low cost disk space and memory these days, we expect that most
users and developers will happily pay this small price for more reliable
software and reduced debugging time. 

Unfortunately, implementing a package manager like NPM for R would require some
fundamental changes in the way R installs and loads packages and namespaces,
which might break backward compatibility at this point. One change that would
probably be required for this is to move away from the \texttt{Depends}
relation definition, and require all packages to rely on \texttt{Imports} and a
\texttt{NAMESPACE} file to explicitly import objects from other packages. A
more challenging problem might be that R should be able to load multiple
versions of a package simultaneously while keeping their namespaces separated.
This is necessary for example when two packages are in use, which
both depend on different versions of one and the same third package. In this
case, the objects, methods and classes exported by the dependency package
should affect only to the package that imported them.

Finally, it would be great if the package manager was capable of installing
multiple versions of a package inside a library, for example by appending
the package version to the name of the installation directory (e.g. \texttt{MASS\_7.3-22}).
The \texttt{library} and \texttt{require} functions could then be extended with
an argument specifying the version to be loaded. This argument could use the
same version range descriptor syntax that packages use to declare dependencies.
Missing versions could automatically be installed, as nothing gets overwritten.

\begin{example}
  library(ggplot2, version="0.8.9")
  library(MASS, version="7.3-x")
  library(Matrix, version=">=1.0")
\end{example}

Code as above leaves little ambiguity and tremendously increases reliability
and reproducibility of R code. When the code is explicit about which package
versions are loaded, and packages are explicit about dependency versions, an R
script or Sweave document that once worked on a certain version of R, will work
for other users, on different systems, and keep working over time, regardless
of upstream changes. For users not concerned with dependency versioning, the
default value of the \texttt{version} argument could be set to \texttt{"*"}.
This value indicates that any version will do, in which case the package
manager gives preference to the most recent available version of the package.

The benefits of a package manager capable of importing specific versions of
packages would not just be limited to contributed code. Such a package
manager would also reduce the necessity to include all of the standard library
and more in the R releases. If implemented, the R Core team could consider
moving some of the \emph{base} and \emph{recommended} packages out of the
\texttt{r-base} distribution, and offer them exclusively through CRAN. This
way, the R software could eventually become the minimal core containing only
the language interpreter and package manager, similar to e.g. Node and NPM.
More high-level functionality could be loaded on demand as versioning is
controlled by the package manager. This would allow for less frequent releases
of the R software itself, and further improve compatibility and reproducibility
between versions of R.

\section{Summary}

The infrastructure for contributed code has supported the steady growth and
adoption of the R software. For the majority of users, contributed code is just
as essential in their daily work as the R base software suite. But the number
of packages on CRAN has grown beyond what could have been foreseen, and
practices and policies that used to work on a smaller scale are becoming
unsustainable. At the same time there is an increasing demand for more
reliable, stable software, that can be used as part of embedded systems,
enterprise applications, or reproducible research. The design and policies of
CRAN and the package manager shape the development process and play an
important role in determining the future of the platform. The current practice
of publishing package updates directly to end-users facilitates a highly
versatile development, but comes at the cost of reliability. The default
behavior of R to install packages in a single library with only the latest
versions is perhaps more appropriate for developers than regular users. After
nearly two decades of development, R has reached a maturity where a slightly
more conservative approach could be beneficial.

This paper explained the problem of dependency versioning, and tried to make
a case for transitioning to a system that does not assume that package versions
are interchangeable. The most straightforward approach would be by extending
the \dfn{r-release} and \dfn{r-devel} branches to the full CRAN repository, and
only publish updates of contributed packages to the \dfn{r-devel}
branch of R. This way, the \dfn{stable} versions of R are tied to a fixed
version of each CRAN package, making the code base and behavior of a
given release of R less ambiguous. Furthermore, a release cycle allows us
to concentrate coordination and testing efforts for contributed packages along
with releases of R, rather than continuously throughout the year.

In the long term, a more fundamental revision of the packaging system could be
considered, in order to facilitate dynamic contributed development without
sacrificing reliability. However, this would involve major changes in the way
libraries and namespaces are managed. The most challenging problem will be
support for concurrently loading multiple versions of a package. 
But when the time is ready to make the jump to the next major release of R, we
hope that R Core will consider revising this important part of the software,
adopting modern approaches and best practices of package management that are
powering collaboration and uniting efforts within other open source communities.

\bibliography{ooms}

\address{Jeroen Ooms\\
  Department of Statistics\\
  University of California\\
  Los Angeles}\\
\email{jeroen.ooms@stat.ucla.edu} \\
\url{http://jeroenooms.github.io/}

\end{article}

\end{document}